\input harvmac
\def\temp{1.35}%
\let\tempp=\relax
\expandafter\ifx\csname psboxversion\endcsname\relax
  \message{PSBOX(\temp)}%
\else
    \ifdim\temp cm>\psboxversion cm
      \message{PSBOX(\temp)}%
    \else
      \message{PSBOX(\psboxversion) is already loaded: I won't load
        PSBOX(\temp)!}%
      \let\temp=\psboxversion
      \let\tempp=\endinput
    \fi
\fi
\tempp
\message{by Jean Orloff: loading ...}
\let\psboxversion=\temp
\catcode`\@=11
%
%
\def\psfortextures{
\def\PSspeci@l##1##2{%
\special{illustration ##1\space scaled ##2}%
}}%
\def\psfordvitops{
\def\PSspeci@l##1##2{%
\special{dvitops: import ##1\space \the\drawingwd \the\drawinght}%
}}%
\def\psfordvips{
\def\PSspeci@l##1##2{%
\d@my=0.1bp \d@mx=\drawingwd \divide\d@mx by\d@my
\includegraphics{##1\space}}}%
\def\psforoztex{
\def\PSspeci@l##1##2{%
\special{##1 \space
      ##2 1000 div dup scale
      \number-\psllx\space\space \number-\pslly\space\space translate
}}}%
\def\psfordvitps{
\def\dvitpsLiter@ldim##1{\dimen0=##1\relax
\special{dvitps: Literal "\number\dimen0\space"}}%
\def\PSspeci@l##1##2{%
\at(0bp;\drawinght){%
\special{dvitps: Include0 "psfig.psr"}
\dvitpsLiter@ldim{\drawingwd}%
\dvitpsLiter@ldim{\drawinght}%
\dvitpsLiter@ldim{\psllx bp}%
\dvitpsLiter@ldim{\pslly bp}%
\dvitpsLiter@ldim{\psurx bp}%
\dvitpsLiter@ldim{\psury bp}%
\special{dvitps: Literal "startTexFig"}%
\special{dvitps: Include1 "##1"}%
\special{dvitps: Literal "endTexFig"}%
}}}%
\def\psfordvialw{
\def\PSspeci@l##1##2{
\special{language "PostScript",
position = "bottom left",
literal "  \psllx\space \pslly\space translate
  ##2 1000 div dup scale
  -\psllx\space -\pslly\space translate",
include "##1"}
}}%
\def\psforptips{
\def\PSspeci@l##1##2{{
\d@mx=\psurx bp
\advance \d@mx by -\psllx bp
\divide \d@mx by 1000\multiply\d@mx by \xscale
\incm{\d@mx}
\let\tmpx\dimincm
\d@my=\psury bp
\advance \d@my by -\pslly bp
\divide \d@my by 1000\multiply\d@my by \xscale
\incm{\d@my}
\let\tmpy\dimincm
\d@mx=-\psllx bp
\divide \d@mx by 1000\multiply\d@mx by \xscale
\d@my=-\pslly bp
\divide \d@my by 1000\multiply\d@my by \xscale
\at(\d@mx;\d@my){\special{ps:##1 x=\tmpx cm, y=\tmpy cm}}
}}}%
\def\psonlyboxes{
\def\PSspeci@l##1##2{%
\at(0cm;0cm){\boxit{\vbox to\drawinght
  {\vss\hbox to\drawingwd{\at(0cm;0cm){\hbox{({\tt##1})}}\hss}}}}
}}%
\def\psloc@lerr#1{%
\let\savedPSspeci@l=\PSspeci@l%
\def\PSspeci@l##1##2{%
\at(0cm;0cm){\boxit{\vbox to\drawinght
  {\vss\hbox to\drawingwd{\at(0cm;0cm){\hbox{({\tt##1}) #1}}\hss}}}}
\let\PSspeci@l=\savedPSspeci@l
}}%
%
%
\newread\pst@mpin
\newdimen\drawinght\newdimen\drawingwd
\newdimen\psxoffset\newdimen\psyoffset
\newbox\drawingBox
\newcount\xscale \newcount\yscale \newdimen\pscm\pscm=1cm
\newdimen\d@mx \newdimen\d@my
\newdimen\pswdincr \newdimen\pshtincr
\let\ps@nnotation=\relax
{\catcode`\|=0 |catcode`|\=12 |catcode`|
|catcode`#=12 |catcode`*=14
|xdef|backslashother{\}*
|xdef|percentother{
|xdef|tildeother{~}*
|xdef|sharpother{#}*
}%
\def\R@moveMeaningHeader#1:->{}%
\def\uncatcode#1{%
\edef#1{\expandafter\R@moveMeaningHeader\meaning#1}}%
\def\execute#1{#1}
\def\psm@keother#1{\catcode`#112\relax}
\def\executeinspecs#1{%
\execute{\begingroup\let\do\psm@keother\dospecials\catcode`\^^M=9#1\endgroup}}%
\def\@mpty{}%
\def\matchexpin#1#2{
  \fi%
  \edef\tmpb{{#2}}%
  \expandafter\makem@tchtmp\tmpb%
  \edef\tmpa{#1}\edef\tmpb{#2}%
  \expandafter\expandafter\expandafter\m@tchtmp\expandafter\tmpa\tmpb\endm@tch%
  \if\match%
}%
\def\matchin#1#2{%
  \fi%
  \makem@tchtmp{#2}%
  \m@tchtmp#1#2\endm@tch%
  \if\match%
}%
\def\makem@tchtmp#1{\def\m@tchtmp##1#1##2\endm@tch{%
  \def\tmpa{##1}\def\tmpb{##2}\let\m@tchtmp=\relax%
  \ifx\tmpb\@mpty\def\match{YN}%
  \else\def\match{YY}\fi%
}}%
\def\incm#1{{\psxoffset=1cm\d@my=#1
 \d@mx=\d@my
  \divide\d@mx by \psxoffset
  \xdef\dimincm{\number\d@mx.}
  \advance\d@my by -\number\d@mx cm
  \multiply\d@my by 100
 \d@mx=\d@my
  \divide\d@mx by \psxoffset
  \edef\dimincm{\dimincm\number\d@mx}
  \advance\d@my by -\number\d@mx cm
  \multiply\d@my by 100
 \d@mx=\d@my
  \divide\d@mx by \psxoffset
  \xdef\dimincm{\dimincm\number\d@mx}
}}%
%
\newif\ifNotB@undingBox
\newhelp\PShelp{Proceed: you'll have a 5cm square blank box instead of
your graphics.}%
\def\s@tsize#1 #2 #3 #4\@ndsize{
  \def\psllx{#1}\def\pslly{#2}%
  \def\psurx{#3}\def\psury{#4}
  \ifx\psurx\@mpty\NotB@undingBoxtrue
  \else
    \drawinght=#4bp\advance\drawinght by-#2bp
    \drawingwd=#3bp\advance\drawingwd by-#1bp
  \fi
  }%
\def\sc@nBBline#1:#2\@ndBBline{\edef\p@rameter{#1}\edef\v@lue{#2}}%
\def\g@bblefirstblank#1#2:{\ifx#1 \else#1\fi#2}%
{\catcode`\%=12
\xdef\B@undingBox{
\def\ReadPSize#1{
 \readfilename#1\relax
 \let\PSfilename=\lastreadfilename
 \openin\pst@mpin=#1\relax
 \ifeof\pst@mpin \errhelp=\PShelp
   \errmessage{I haven't found your postscript file (\PSfilename)}%
   \psloc@lerr{was not found}%
   \s@tsize 0 0 142 142\@ndsize
   \closein\pst@mpin
 \else
   \if\matchexpin{\GlobalInputList}{, \lastreadfilename}%
   \else\xdef\GlobalInputList{\GlobalInputList, \lastreadfilename}%
     \immediate\write\psbj@inaux{\lastreadfilename,}%
   \fi%
   \loop
     \executeinspecs{\catcode`\ =10\global\read\pst@mpin to\n@xtline}%
     \ifeof\pst@mpin
       \errhelp=\PShelp
       \errmessage{(\PSfilename) is not an Encapsulated PostScript File:
           I could not find any \B@undingBox: line.}%
       \edef\v@lue{0 0 142 142:}%
       \psloc@lerr{is not an EPSFile}%
       \NotB@undingBoxfalse
     \else
       \expandafter\sc@nBBline\n@xtline:\@ndBBline
       \ifx\p@rameter\B@undingBox\NotB@undingBoxfalse
         \edef\t@mp{%
           \expandafter\g@bblefirstblank\v@lue\space\space\space}%
         \expandafter\s@tsize\t@mp\@ndsize
       \else\NotB@undingBoxtrue
       \fi
     \fi
   \ifNotB@undingBox\repeat
   \closein\pst@mpin
 \fi
\message{#1}%
}%
%
%
\def\psboxto(#1;#2)#3{\vbox{%
   \ReadPSize{#3}%
   \advance\pswdincr by \drawingwd
   \advance\pshtincr by \drawinght
   \divide\pswdincr by 1000
   \divide\pshtincr by 1000
   \d@mx=#1
   \ifdim\d@mx=0pt\xscale=1000
         \else \xscale=\d@mx \divide \xscale by \pswdincr\fi
   \d@my=#2
   \ifdim\d@my=0pt\yscale=1000
         \else \yscale=\d@my \divide \yscale by \pshtincr\fi
   \ifnum\yscale=1000
         \else\ifnum\xscale=1000\xscale=\yscale
                    \else\ifnum\yscale<\xscale\xscale=\yscale\fi
              \fi
   \fi
   \divide\drawingwd by1000 \multiply\drawingwd by\xscale
   \divide\drawinght by1000 \multiply\drawinght by\xscale
   \divide\psxoffset by1000 \multiply\psxoffset by\xscale
   \divide\psyoffset by1000 \multiply\psyoffset by\xscale
   \global\divide\pscm by 1000
   \global\multiply\pscm by\xscale
   \multiply\pswdincr by\xscale \multiply\pshtincr by\xscale
   \ifdim\d@mx=0pt\d@mx=\pswdincr\fi
   \ifdim\d@my=0pt\d@my=\pshtincr\fi
   \message{scaled \the\xscale}%
 \hbox to\d@mx{\hss\vbox to\d@my{\vss
   \global\setbox\drawingBox=\hbox to 0pt{\kern\psxoffset\vbox to 0pt{%
      \kern-\psyoffset
      \PSspeci@l{\PSfilename}{\the\xscale}%
      \vss}\hss\ps@nnotation}%
   \global\wd\drawingBox=\the\pswdincr
   \global\ht\drawingBox=\the\pshtincr
   \global\drawingwd=\pswdincr
   \global\drawinght=\pshtincr
   \baselineskip=0pt
   \copy\drawingBox
 \vss}\hss}%
  \global\psxoffset=0pt
  \global\psyoffset=0pt
  \global\pswdincr=0pt
  \global\pshtincr=0pt 
  \global\pscm=1cm 
}}%
%
%
\def\psboxscaled#1#2{\vbox{%
  \ReadPSize{#2}%
  \xscale=#1
  \message{scaled \the\xscale}%
  \divide\pswdincr by 1000 \multiply\pswdincr by \xscale
  \divide\pshtincr by 1000 \multiply\pshtincr by \xscale
  \divide\psxoffset by1000 \multiply\psxoffset by\xscale
  \divide\psyoffset by1000 \multiply\psyoffset by\xscale
  \divide\drawingwd by1000 \multiply\drawingwd by\xscale
  \divide\drawinght by1000 \multiply\drawinght by\xscale
  \global\divide\pscm by 1000
  \global\multiply\pscm by\xscale
  \global\setbox\drawingBox=\hbox to 0pt{\kern\psxoffset\vbox to 0pt{%
     \kern-\psyoffset
     \PSspeci@l{\PSfilename}{\the\xscale}%
     \vss}\hss\ps@nnotation}%
  \advance\pswdincr by \drawingwd
  \advance\pshtincr by \drawinght
  \global\wd\drawingBox=\the\pswdincr
  \global\ht\drawingBox=\the\pshtincr
  \global\drawingwd=\pswdincr
  \global\drawinght=\pshtincr
  \baselineskip=0pt
  \copy\drawingBox
  \global\psxoffset=0pt
  \global\psyoffset=0pt
  \global\pswdincr=0pt
  \global\pshtincr=0pt 
  \global\pscm=1cm
}}%
%
\def\psbox#1{\psboxscaled{1000}{#1}}%
\newif\ifn@teof\n@teoftrue
\newif\ifc@ntrolline
\newif\ifmatch
\newread\j@insplitin
\newwrite\j@insplitout
\newwrite\psbj@inaux
\immediate\openout\psbj@inaux=psbjoin.aux
\immediate\write\psbj@inaux{\string\joinfiles}%
\immediate\write\psbj@inaux{\jobname,}%
%
%
\def\toother#1{\ifcat\relax#1\else\expandafter%
  \toother@ux\meaning#1\endtoother@ux\fi}%
\def\toother@ux#1 #2#3\endtoother@ux{\def\tmp{#3}%
  \ifx\tmp\@mpty\def\tmp{#2}\let\next=\relax%
  \else\def\next{\toother@ux#2#3\endtoother@ux}\fi%
\next}%
%
%
\let\readfilenamehook=\relax
\def\re@d{\expandafter\re@daux}
\def\re@daux{\futurelet\nextchar\stopre@dtest}%
\def\re@dnext{\xdef\lastreadfilename{\lastreadfilename\nextchar}%
  \afterassignment\re@d\let\nextchar}%
\def\stopre@d{\egroup\readfilenamehook}%
\def\stopre@dtest{%
  \ifcat\nextchar\relax\let\nextread\stopre@d
  \else
    \ifcat\nextchar\space\def\nextread{%
      \afterassignment\stopre@d\chardef\nextchar=`}%
    \else\let\nextread=\re@dnext
      \toother\nextchar
      \edef\nextchar{\tmp}%
    \fi
  \fi\nextread}%
\def\readfilename{\bgroup%
  \let\\=\backslashother \let\%=\percentother \let\~=\tildeother
  \let\#=\sharpother \xdef\lastreadfilename{}%
  \re@d}%
%
%
\xdef\GlobalInputList{\jobname}%
\def\psnewinput{%
  \def\readfilenamehook{
    \if\matchexpin{\GlobalInputList}{, \lastreadfilename}%
    \else\xdef\GlobalInputList{\GlobalInputList, \lastreadfilename}%
      \immediate\write\psbj@inaux{\lastreadfilename,}%
    \fi%
    \let\readfilenamehook=\relax%
    \ps@ldinput\lastreadfilename\relax%
  }\readfilename%
}%
\expandafter\ifx\csname @@input\endcsname\relax    
  \immediate\let\ps@ldinput=\input\def\input{\psnewinput}%
\else
  \immediate\let\ps@ldinput=\@@input
  \def\@@input{\psnewinput}%
\fi%
\def\nowarnopenout{%
 \def\warnopenout##1##2{%
   \readfilename##2\relax
   \message{\lastreadfilename}%
   \immediate\openout##1=\lastreadfilename\relax}}%
\def\warnopenout#1#2{%
 \readfilename#2\relax
 \def\t@mp{TrashMe,psbjoin.aux,psbjoint.tex,}\uncatcode\t@mp
 \if\matchexpin{\t@mp}{\lastreadfilename,}%
 \else
   \immediate\openin\pst@mpin=\lastreadfilename\relax
   \ifeof\pst@mpin
     \else
     \edef\tmp{{If the content of this file is precious to you, this
is your last chance to abort (ie press x or e) and rename it before
retexing (\jobname). If you're sure there's no file
(\lastreadfilename) in the directory of (\jobname), then go on: I'm
simply worried because you have another (\lastreadfilename) in some
directory I'm looking in for inputs...}}%
     \errhelp=\tmp
     \errmessage{I may be about to replace your file named \lastreadfilename}%
   \fi
   \immediate\closein\pst@mpin
 \fi
 \message{\lastreadfilename}%
 \immediate\openout#1=\lastreadfilename\relax}%
{\catcode`\%=12\catcode`\*=14
\gdef\splitfile#1{*
 \readfilename#1\relax
 \immediate\openin\j@insplitin=\lastreadfilename\relax
 \ifeof\j@insplitin
   \message{! I couldn't find and split \lastreadfilename!}*
 \else
   \immediate\openout\j@insplitout=TrashMe
   \message{< Splitting \lastreadfilename\space into}*
   \loop
     \ifeof\j@insplitin
       \immediate\closein\j@insplitin\n@teoffalse
     \else
       \n@teoftrue
       \executeinspecs{\global\read\j@insplitin to\spl@tinline\expandafter
         \ch@ckbeginnewfile\spl@tinline
       \ifc@ntrolline
       \else
         \toks0=\expandafter{\spl@tinline}*
         \immediate\write\j@insplitout{\the\toks0}*
       \fi
     \fi
   \ifn@teof\repeat
   \immediate\closeout\j@insplitout
 \fi\message{>}*
}*
\gdef\ch@ckbeginnewfile#1
 \def\t@mp{#1}*
 \ifx\@mpty\t@mp
   \def\t@mp{#3}*
   \ifx\@mpty\t@mp
     \global\c@ntrollinefalse
   \else
     \immediate\closeout\j@insplitout
     \warnopenout\j@insplitout{#2}*
     \global\c@ntrollinetrue
   \fi
 \else
   \global\c@ntrollinefalse
 \fi}*
\gdef\joinfiles#1\into#2{*
 \message{< Joining following files into}*
 \warnopenout\j@insplitout{#2}*
 \message{:}*
 {*
 \edef\w@##1{\immediate\write\j@insplitout{##1}}*
\w@{
\w@{
\w@{
\w@{
\w@{
\w@{
\w@{
\w@{
\w@{
\w@{
\w@{\string\input\space psbox.tex}*
\w@{\string\splitfile{\string\jobname}}*
\w@{\string\let\string\autojoin=\string\relax}*
}*
 \expandafter\tre@tfilelist#1, \endtre@t
 \immediate\closeout\j@insplitout
 \message{>}*
}*
\gdef\tre@tfilelist#1, #2\endtre@t{*
 \readfilename#1\relax
 \ifx\@mpty\lastreadfilename
 \else
   \immediate\openin\j@insplitin=\lastreadfilename\relax
   \ifeof\j@insplitin
     \errmessage{I couldn't find file \lastreadfilename}*
   \else
     \message{\lastreadfilename}*
     \immediate\write\j@insplitout{
     \executeinspecs{\global\read\j@insplitin to\oldj@ininline}*
     \loop
       \ifeof\j@insplitin\immediate\closein\j@insplitin\n@teoffalse
       \else\n@teoftrue
         \executeinspecs{\global\read\j@insplitin to\j@ininline}*
         \toks0=\expandafter{\oldj@ininline}*
         \let\oldj@ininline=\j@ininline
         \immediate\write\j@insplitout{\the\toks0}*
       \fi
     \ifn@teof
     \repeat
   \immediate\closein\j@insplitin
   \fi
   \tre@tfilelist#2, \endtre@t
 \fi}*
}%
\def\autojoin{%
 \immediate\write\psbj@inaux{\string\into{psbjoint.tex}}%
 \immediate\closeout\psbj@inaux
 \expandafter\joinfiles\GlobalInputList\into{psbjoint.tex}%
}%
%
%
%
\def\centinsert#1{\midinsert\line{\hss#1\hss}\endinsert}%
\def\psannotate#1#2{\vbox{%
  \def\ps@nnotation{#2\global\let\ps@nnotation=\relax}#1}}%
\def\pscaption#1#2{\vbox{%
   \setbox\drawingBox=#1
   \copy\drawingBox
   \vskip\baselineskip
   \vbox{\hsize=\wd\drawingBox\setbox0=\hbox{#2}%
     \ifdim\wd0>\hsize
       \noindent\unhbox0\tolerance=5000
    \else\centerline{\box0}%
    \fi
}}}%
%
\def\at(#1;#2)#3{\setbox0=\hbox{#3}\ht0=0pt\dp0=0pt
  \rlap{\kern#1\vbox to0pt{\kern-#2\box0\vss}}}%
%
\newdimen\gridht \newdimen\gridwd
\def\gridfill(#1;#2){%
  \setbox0=\hbox to 1\pscm
  {\vrule height1\pscm width.4pt\leaders\hrule\hfill}%
  \gridht=#1
  \divide\gridht by \ht0
  \multiply\gridht by \ht0
  \gridwd=#2
  \divide\gridwd by \wd0
  \multiply\gridwd by \wd0
  \advance \gridwd by \wd0
  \vbox to \gridht{\leaders\hbox to\gridwd{\leaders\box0\hfill}\vfill}}%
%
\def\fillinggrid{\at(0cm;0cm){\vbox{%
  \gridfill(\drawinght;\drawingwd)}}}%
%
%
\def\textleftof#1:{%
  \setbox1=#1
  \setbox0=\vbox\bgroup
    \advance\hsize by -\wd1 \advance\hsize by -2em}%
\def\textrightof#1:{%
  \setbox0=#1
  \setbox1=\vbox\bgroup
    \advance\hsize by -\wd0 \advance\hsize by -2em}%
\def\endtext{%
  \egroup
  \hbox to \hsize{\valign{\vfil##\vfil\cr%
\box0\cr%
\noalign{\hss}\box1\cr}}}%
%
\def\frameit#1#2#3{\hbox{\vrule width#1\vbox{%
  \hrule height#1\vskip#2\hbox{\hskip#2\vbox{#3}\hskip#2}%
        \vskip#2\hrule height#1}\vrule width#1}}%
\def\boxit#1{\frameit{0.4pt}{0pt}{#1}}%
\catcode`\@=12 
%
\psfordvips   
\let\fillinggrid=\relax      
\baselineskip=12pt
\overfullrule=0pt
\def\nl{\hfil\break}
\def\half{{1\over2}}
\def\c#1{{\cal #1}}
\def\IR{\relax{\rm I\kern-.18em R}}
\def\endpage{\vfil\eject}
\def\corru#1#2{\langle #2\rangle_{{}_{#1}}}
\def\d{\partial}
\def\prodone#1{\prod_{j=1}^{#1}}
\def\del#1{\partial_{#1}}
\def\Re{{\rm Re}}
\def\Im{{\rm Im}}
\def\pint{{\int \!\!\!\!\!\! -}}
\def\al#1{a_{#1}}
\def\vsiz{2.0cm}
\def\em#1{\it #1}
\def\qq#1{q_{#1}}
\def\section#1{\newsec{#1}}
\def\anp{{ \sl Ann. Phys. }}
\def\cmp{{ \sl Comm. Math. Phys. }}
\def\ijp{{ \sl Int. J. of Mod. Phys. }}
\def\jpa{{ \sl J. of Phys. }}
\def\lmp{{ \sl Lett. Math. Phys. }}
\def\mpl{{ \sl Mod. Phys. Lett. }}
\def\nvc{{ \sl Nuovo Cim. }}
\def\npb{{ \sl Nucl. Phys. }}
\def\prc{{ \sl Phys. Rep. }}
\def\prd{{ \sl Phys. Rev. }}
\def\prl{{ \sl Phys. Rev. Lett. }}
\def\plb{{ \sl Phys. Lett. }}
\def\rmp{{ \sl Rev. Mod. Phys. }}
\def\sovjnp{{ \sl Sov. J. of Nucl. Phys. }}
\def\tmp{{ \sl Theor. Math. Phys. }}
\def\zpc{{ \sl Zeit. Phys. }}
\def\tpl{\tilde\varphi_1}
\def\tpr{\tilde\varphi_2}
\def\ct{\cos\gamma}
\def\st{\sin\gamma}
\def\cst{\cos^2\gamma}
\def\sst{\sin^2\gamma}
\def\dk{\int\!{d^4k\over(2\pi)^4}\,}
\def\mo{{m_1^2}}
\def\mt{{m_2^2}}
\def\ml{{m_L^2}}
\def\kmn#1#2{K_{\mu\nu}(p^2;#1,#2)}
\def\IZ{\relax\ifmmode\mathchoice
  {\hbox{\cmss Z\kern-.4em Z}}{\hbox{\cmss Z\kern-.4em Z}}
  {\lower.9pt\hbox{\cmsss Z\kern-.4em Z}}
  {\lower1.2pt\hbox{\cmsss Z\kern-.4em Z}}\else{\cmss Z\kern-.4em Z}\fi}
\Title{}{
 Non--decoupling effects due to a dimensionful coupling
}
\centerline{{
Kenichiro Aoki}\footnote{$^1$}
{Email: {\tt~ken@phys-h.keio.ac.jp}}
}
\vskip.8cm
\centerline{\it Department of Physics, Hiyoshi Campus} 
 \centerline{\it Keio University}
  \centerline{\it 4--1--1 Hiyoshi, Kouhoku-ku}
  \centerline{\it Yokohama { 223},   JAPAN}
\vskip .4cm
\centerline{{\bf Abstract}}
We show that the dimensionful scalar cubic coupling in $3+1$
dimensions gives rise to non--decoupling effects and analyze
the behavior of these effects.  
In the process, we clarify how it is perturbatively consistent
to construct theories in which the cubic coupling is dominant.
We use the scalar sector of the
Supersymmetric Standard Model as an example.
We discuss how the non--decoupling effects may be analyzed
systematically. 
\bigskip

\Date{}
\vfill\eject
\section{Introduction}
The properties of particles too heavy to be produced in
experiments may sometimes be extracted through their physically
observable quantum effects.  Under reasonably general
conditions, the decoupling theorem \ref\AC{T.~Appelquist,
  J.~Carazzone, \prd{\bf11} (1975) 2856} applies and these
effects become smaller as their masses become larger.  However,
there also exist so--called ``non--decoupling effects'' wherein
the assumptions of the decoupling theorem are {\it not}
satisfied.  In such a situation, a heavy particle can give rise
to physically observable quantum effects that do not necessarily
decrease with the particle mass.  The importance of these
non--decoupling effects hardly needs to be stressed; they were
used to obtain a perturbative bound for the top mass
\ref\TOPBOUND{ Review of particle properties, \prd{\bf D45}
  (1992) S1, \plb{\bf 239B} (1990) 1 and references therein. }
which was borne out in
experiment\ref\FNAL{CDF Collaboration (F. Abe et al.), \prl{\bf
    74} (1995) 2626\nl D0 Collaboration (S. Abachi et al.),
  \prl{\bf 74} (1995) 2632}.  (Non--perturbative considerations
are needed for a very heavy top mass, which have been analyzed
in \ref\APRHO{ K.~Aoki, S.~Peris,   \zpc{\bf C61} (1994)
  303\nl
  K.~Aoki, S.~Peris,   \plb{\bf B335} (1994) 470
  }%
\ref\APS{K.~Aoki, S.~Peris, \prl{\bf 70} (1993) 1743}%
\ref\CPP{S. Cortese,  E. Pallante, R. Petronzio, \plb{\bf 
    301B} (1993) 203 }.)  
Non--decoupling effects have also been
instrumental in the investigations of observable effects of the
various extensions of the Standard Model, which have been
performed at an impressive level. 
\ref\PDG{ Review of particle properties, \prd{\bf D54}
  (1996) 1, and references therein. }

While the importance of non--decoupling effects is universally
accepted, the properties of non--decoupling effects are not
systematically understood.  This is to be contrasted to the
decoupling effects which are known to decrease as a power of the
particle mass, up to logarithms, when we increase the particle
mass as we keep the coupling constant fixed \AC.  To
systematically understand non--decoupling effects, we first need
to establish which couplings may give rise to non--decoupling
effects.  Furthermore, we need to analyze the mass dependence of
these effects since it is of crucial importance in determining
the actual size of the effects.  It is desirable to have a
systematic understanding of these effects, wherein the couplings
are classified as those that may give rise to non--decoupling
effects or those which do not, and if they do give rise to
non--decoupling effects, what their possible leading order mass
dependence may be.  Such a systematic understanding, even
perturbatively, is lacking at this point.  

It is well--known that the dimensionless couplings, namely the
Yukawa and the scalar quartic coupling, give rise to
non--decoupling effects.  For instance, in the Standard Model,
the top and the Higgs contribute to the $\rho$ parameter at the
one loop level as
\eqn\drhosm{\delta\rho_{\rm top}={3\over (4\pi)^2} \sqrt 2
  G_Fm_t^2 \quad,
  \qquad\qquad
  \delta\rho_{\rm Higgs}=
  - {3\over 4} {g'^2\over (4\pi)^2}\log ({M_H^2\over M_W^2})
  }
Quantum corrections due to gauge couplings present a more
complicated case which we shall not deal with here.
On the other hand, we know that the terms up to quadratic order
with respect to the fields in the action --- the kinetic and the
mass terms --- cannot give rise to non--decoupling effects by
themselves since they do not generate interactions.  To be
systematic, this leaves the case of the dimensionful scalar
cubic coupling within the renormalizable couplings in four
dimensions.  This case is of interest, since in some sense it
lies in between the couplings that are known to give rise to
non--decoupling effects and those that do not.  Also, the {\it
  dimensionless} property of the couplings that give rise to
non--decoupling effects is sometimes stressed \ref\JC{See for
  example, J.C.  Collins, {``\sl Renormalization'',} Cambridge
  University Press (1984), Chapter 8.}.  It is of interest to
find if the dimensionful couplings give rise to non--decoupling
effects and if so, how the behavior non--decoupling effects
changes with the dimensionality of the coupling.  Furthermore,
these coupling arise naturally in some popular extensions of the
Standard Model, such as the Supersymmetric Standard Model and it
is important to understand what kind of physical effects the
coupling produces.

In this note, we shall show that the cubic scalar coupling by
itself 
generically {\it does} give rise to non--decoupling effects
perturbatively and analyze the mass dependence of these effects.
We find that the behavior is qualitatively different from those
previously analyzed for the other couplings.
While the need for
investigating the effects of the scalar cubic coupling seems
obvious, there is perhaps a reason why it has not been
investigated { by itself} previously;  the scalar cubic
coupling tends to give rise to a classical instability in the
absence of quartic couplings.  However, this needs not be
the case, as we shall see below.
\newsec{The model}
The model we consider has the following Lagrangian
\eqn\lag{
  -\c L=\left|\del\mu\phi\right|^2
  +\left|\del\mu\varphi_L\right|^2 
  +m_L^2\left|\varphi_L\right|^2
  +\left|\del\mu\varphi_R\right|^2
  +m_R^2\left|\varphi_R\right|^2
  +\lambda
  \left(\varphi_L^\dagger \phi\varphi_R+
  \varphi_R^\dagger\phi^\dagger\varphi_L\right)
  } 
$\phi,\varphi_L$ are scalar doublets and $\varphi_R$ is a scalar
singlet; the reason for this choice of fields will become clear
below. The model has {\it no} classical instability and has a
flat direction along the $\phi$ direction.  The symmetry of this
model is $U(2)_L\times U(1)_R$ that acts on the fields of the
theory thus
\eqn\symaction{
  \varphi_L\mapsto U_L\varphi_L\qquad
  \varphi_R\mapsto e^{i\alpha}\varphi_R\qquad
  \phi\mapsto U_L\phi e^{-i\alpha}\qquad
  U\in {\rm U}(2),\ \alpha\in\IR
  } 
Some of the quartic couplings not included in the action allowed
by symmetry are generated at the quantum level.
Perturbatively, these quartic terms will have couplings which
are of one loop order which only contribute at the two loop
level and we shall not consider them below. In other words, we
consider the region where the scalar cubic coupling is strong
compared to other couplings and it is consistent to do so within
perturbation theory.  We shall elaborate more on this below.

This model may seem somewhat contrived.  However, this model
arises in the Supersymmetric Standard Model, wherein the scalar
cubic coupling is the soft supersymmetry breaking parameter.
There are additional couplings in the Supersymmetric Standard
model, namely the electroweak gauge couplings --- which are, of
course weak --- and the cubic {\it superpotential} (Yukawa)
coupling, that both give rise to quartic some scalar couplings.
The Yukawa coupling is weak for light fermion multiplets. In the
Supersymmetric Standard Model, $\phi,\varphi_L$ and $\varphi_R$
correspond to one of the Higgs doublets, the supersymmetric
partner of the fermion $SU(2)_L$ doublet and the singlet,
respectively.  We have chosen a model that is closely related to
the Standard Model since it is relatively simple and it allows
us to compute physical parameters with physical significance
familiar to us.  While the effects of the scalar cubic coupling
has been included in studies of the Supersymmetric Standard
Model, the effect of the cubic coupling {\it by itself } has not
been cleanly delineated, to our knowledge.

In the Standard Model, the model is gauged by changing the
derivatives to the covariant derivatives.   The covariant
derivatives act on the fields as 
\eqn\covderivs{
  \eqalign{
  D_\mu\varphi_L&\equiv
  \left(\d_\mu-ig W_\mu^a {\sigma^a\over2}
    -ig'{Y_L\over2}B_\mu\right)\varphi_L,\quad
  D_\mu\varphi_R\equiv
  \left(\d_\mu-ig'{Y_R\over2}B_\mu\right)\varphi_R,\cr
  D_\mu\phi&\equiv
  \left(\d_\mu-ig W_\mu^a {\sigma^a\over2}
    +i{g'\over2}B_\mu\right)\phi,\qquad Y_R-Y_L=1\cr}
}  
The gauge couplings are considered as being weak in this model
and quantum corrections due to the gauge sector will not be
considered below.

The $\phi$ field has the vacuum expectation value along the flat
direction,
which breaks the symmetry to $U(1)\times U(1)$.  
\eqn\phivev{
  \langle\phi\rangle=\pmatrix{v/\sqrt2\cr0\cr},\qquad
  \phi\equiv\pmatrix{1/\sqrt2(v+\sigma+\chi^0)\cr\chi^-\cr}.
  }
If desired, this may be uniquely arranged by including quartic
couplings in the zero coupling limit analogous to the
Prasad--Sommerfeld limit or using weak quartic couplings as in
the Supersymmetric Standard Model.
More concretely, we may consider adding a quartic interaction
term $h (\phi^\dagger\phi-v^2/2)^2$.  Since this term only has
to stabilize the vacuum expectation value against radiative
corrections, the size of the coupling $h$ needs to be of only
one loop order.  Therefore, its contribution to the physical
parameters is at the two loop level within perturbation theory.
This statement may be made more precise if necessary: We could
easily generalize the theory by replacing SU$(2)_L$ by
SU$(N)_L$.  In this case, we may control the size of the
couplings systematically by using the large $N$ expansion.  If
we let $\lambda$ be of order $N^{-1/2}$ in the standard fashion,
the corrections to the $\phi^\dagger\phi$, $(\phi^\dagger\phi)^2
$ terms are of order $\c O(N^{-1})$, $\c O(N^{-2})$ respectively
so that the flat direction is preserved to leading order.
Therefore, it is consistent to set a vacuum expectation value
for $\phi$ as in \phivev. If desired, we may put in a coupling
$h$ of order $N^{-b}\ (1<b<2) $, which does not produce any
leading order radiative corrections but is enough to stabilize
the vacuum.

When $v\not=0$, the $\left(\varphi_L\right)_1 $ and $\varphi_R$ 
are no longer mass eigenstates.  The mass eigenstates
$\tpl,\tpr$ are
\eqn\masseigenstates{
  \pmatrix{\left(\varphi_L\right)_1\cr \varphi_R\cr}
  =
  \pmatrix{\ct & -\st\cr \st&\ct\cr}
  \pmatrix{\tpl\cr\tpr\cr}  
  }
where
\eqn\thetadef{
  \tan\gamma={1\over\sqrt2\lambda v}
  \left[-m_L^2+m_R^2 + \sqrt{
      (m_L^2-m_R^2)^2+2\lambda^2v^2}\right]
  }
The masses of $\tpl,\tpr$ are 
\eqn\masses{
  m_1^2, m_2^2= \half\left[m_L^2+m_R^2
    \pm\sqrt{
      (m_L^2-m_R^2)^2+2\lambda^2v^2}\right]
  }
The mass parameters satisfy the inequalities 
$  0\leq m_1^2\leq m_L^2 \leq m_2^2$
and are otherwise independent.
The particles in this model and their masses are
  $\phi\ [0],\ \tpl\ [\ m_1],\ 
  \tpr\ [\ m_2],\ 
  \left(\varphi_L\right)_2\ [m_L]$.
\section{Contribution to the $\rho$ parameter}
To investigate the possibility of non--decoupling effects, we
need to study a physically observable effect that is generated
by quantum effects.  Here, we first choose to study a
well--known physical parameter; the contribution of this model
to the so called $\rho$ parameter,
$\rho=\left.{M_W^2/
      (M_Z^2\cos^2\theta_W)}\right|_{p^2=0}$
\ref\RHO{M.~Veltman, \npb{\bf B123} (1977) 89}
This parameter measures the asymmetry between the charged and
the neutral gauge interactions at low energies, measured for
instance in the ratio of low energy neutrino cross sections.  We
will obtain this contribution by two methods, first via the
corrections to the scalar kinetic term and also through the
current correlation functions.  We show that they both give rise
to the same contribution to leading order in the weak couplings,
as they should.
\subsec{Quantum corrections to the scalar kinetic terms}
The contribution to $\rho$ is
\eqn\drhoz{
  \rho= \left.{Z_+\over Z_0}\right|_{\rm zero\  momentum}=
  1+\left(Z_+-Z_0\right)_{\rm zero\ momentum}+\c O(Z-1)^2
  }
where $Z_+,Z_0$ are the coefficients of the kinetic terms for
$\chi^0,\chi^+$. 
The $\rho$ parameter may be
obtained from the ungauged scalar theory for the following
reason
\ref\LYTEL{R.S.~Lytel, \prd{\bf 22D} (1980) 505
\nl A.C.~Longhitano, \npb{\bf B188} (1981) 118}%
\ref\PP{R.D. Peccei, S.~Peris  \prd{\bf D44} (1991) 809}%
\ref\FMM{F. Feruglio, A. Masiero, L. Maiani, \npb{\bf B387}
  (1992) 523}%
: The $\rho$ parameter may be thought of as the ratio of
the masses of the gauge bosons.  The mass terms for the gauge
bosons may be obtained by minimally gauging the Nambu--Goldstone
scalar kinetic terms.  The coefficient for this kinetic term can
hence be obtained within the scalar theory.  This may be
illustrated in the effective action if we consider the lowest
dimension operator that contributes to the $\rho$ parameter; in
this case, there is a unique operator that contributes to the
scalar kinetic term, $\left(\phi^\dagger D_\mu\phi\right)^2$.
Therefore, it may be obtained solely from the scalar theory
without gauge interactions, to leading order.

To obtain the $Z$ factors, we compute the vacuum polarization
graphs in
\fig\figrhong{}.
\nl\nl\centerline{
  \psannotate{\psboxto(0cm;\vsiz){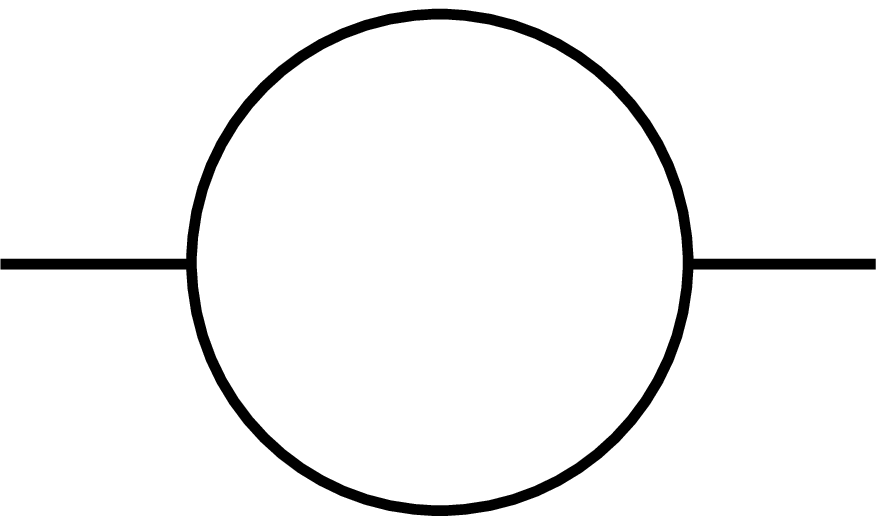}}
  {\fillinggrid%
    \at(0.5\pscm;3.0\pscm){$\chi^0$}%
    \at(8.0\pscm;3.0\pscm){$\chi^0$}%
    \at(6.0\pscm;5.0\pscm){$\tpl$}%
    \at(2.0\pscm;0.0\pscm){$\tpr$}%
    }
  \hskip1.2cm
  \psannotate{\psboxto(0cm;\vsiz){vscal.eps}}
  {\fillinggrid%
    \at(0.5\pscm;3.0\pscm){$\chi^+$}%
    \at(8.0\pscm;3.0\pscm){$\chi^+$}%
    \at(6.0\pscm;5.0\pscm){$\tpl,\tpr$}%
    \at(1.5\pscm;-0.2\pscm){$\left(\varphi_L\right)_2$}%
    }
    } \nl
\figrhong\ {\it %
  Contributions to the $\rho$ parameter computed within the
  scalar
  theory. %
}\nl\nl
and obtain the self energies
\eqn\pifactors{
    \pi_0(p^2)=\lambda^2 I_0(p^2;m_1^2,m_2^2)
    ,\quad
    \pi_+(p^2)=\lambda^2 \left[\sst\, I_0(p^2;m_1^2,m_L^2)+
      \cst\, I_0(p^2;m_2^2,m_L^2)\right]
  }
where we defined 
\eqn\izdef{
  I_0(p^2;m_a^2,m_b^2)\equiv
  \dk {1\over\left((k+p)^2+m_a^2\right)\left(k^2+m_b^2\right)}
  }
After some manipulation, using $Z=1-d\pi/dp^2$,  we obtain a
compact formula
\eqn\drhofinal{
  \delta\rho=
  -{\lambda^4v^2 \over 32\pi^2}\int_0^\infty\!\!dx\,
  {x\left(x^2-m_1^2m_2^2\right)
    \over\left(x+m_1^2\right)^3
    \left(x+m_2^2\right)^3
    \left(x+m_L^2\right)}
  }
This expression is finite both in the infrared and ultraviolet,
as it should be.  We explained how the correction corresponds to
the $\delta\rho$ above.  We stress here that this parameter is a
physical parameter within the {\it ungauged } scalar theory; it
describes the quantum generated asymmetry between the scalars at
low energies, realized, for instance in scattering.
\subsec{Current correlation functions}
The $\rho$ parameter is the ratio of the strength of the charged
current interactions to the neutral current interactions in the
low energy limit.  The correction to the $\rho$ parameter is
\eqn\rhodefgauge{\rho=\left.{M_W^2\over M_Z^2\cos^2\theta_W}\right|_
  {p^2=0}
  = 1- {\Pi_W\over M_W^2}+{\Pi_Z\over M_Z^2} +{\rm higher \ orders.}
  }
Here we defined $\Pi_{W,Z}$ to be the
the vacuum polarization of the gauge bosons.
All the quantities in the above expression are evaluated at zero 
momentum.
We compute the following current correlation functions
\fig\figrhow{}
\nl\nl\centerline{
  \psannotate{\psboxto(0cm;\vsiz){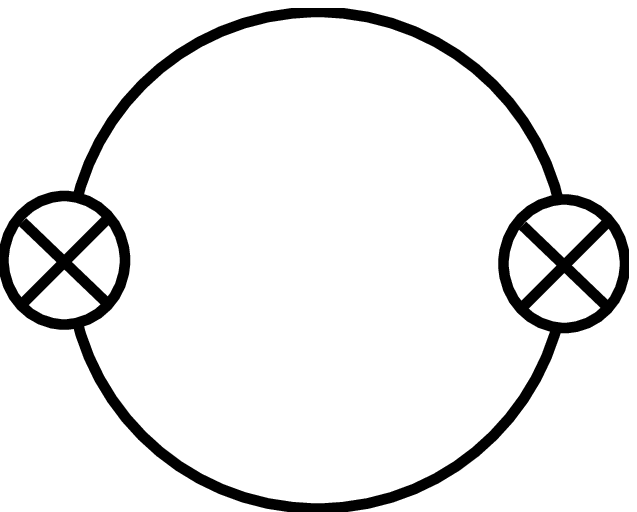}}
  {\fillinggrid%
    \at(-0.7\pscm;3.0\pscm){$Z$}%
    \at(6.5\pscm;3.0\pscm){$Z$}%
    \at(5.0\pscm;5.0\pscm){$\tpl$}%
    \at(0.8\pscm;0.0\pscm){$\tpl$}%
    }
  \hskip1.2cm
  \psannotate{\psboxto(0cm;\vsiz){vcurr.eps}}
  {\fillinggrid%
    \at(-0.7\pscm;3.0\pscm){$Z$}%
    \at(6.5\pscm;3.0\pscm){$Z$}%
    \at(5.0\pscm;5.0\pscm){$\tpr$}%
    \at(0.8\pscm;0.0\pscm){$\tpr$}%
    }
  \hskip1.2cm
  \psannotate{\psboxto(0cm;\vsiz){vcurr.eps}}
  {\fillinggrid%
    \at(-0.7\pscm;3.0\pscm){$Z$}%
    \at(6.5\pscm;3.0\pscm){$Z$}%
    \at(5.0\pscm;5.1\pscm){$\left(\varphi_L\right)_2$}%
    \at(0.0\pscm;-0.2\pscm){$\left(\varphi_L\right)_2$}%
    }
   \hskip1.2cm
  \psannotate{\psboxto(0cm;\vsiz){vcurr.eps}}
  {\fillinggrid%
    \at(-1.1\pscm;3.0\pscm){$W^+$}%
    \at(6.5\pscm;3.0\pscm){$W^+$}%
    \at(5.0\pscm;5.0\pscm){$\tpl,\tpr$}%
    \at(0.0\pscm;-0.2\pscm){$\left(\varphi_L\right)_2$}%
    }
  } \nl
\figrhow\ {\it %
Contributions to the current correlation functions needed for
obtaining the $\rho$ parameter. 
}\nl\nl
to obtain
\eqn\currents{
  \eqalign{
    \Pi_Z(p^2=0)&={g^2\over4}\left[\cos^4\gamma\, I_1(m_1^2,m_1^2)+
      \sin^4\gamma\, I_1(m_2^2,m_2^2) + I_1(m_L^2,m_L^2)
      +2\sst\cst \,I_1(m_1^2,m_2^2)\right]\cr
    \Pi_W(p^2=0) &= {g^2\over2}\left[\cst\, I_1(m_1^2,m_L^2)
      +\sst\, I_1(m_2^2,m_L^2)\right]\cr}
  }
where we defined 
The seagull contributions cancel out in the expression for
$\delta\rho$ and will not be discussed below.
We obtain another compact formula for $\delta\rho$
\eqn\drhow{\delta\rho
  ={\lambda^4v^2\over64\pi^2}
  \int_0^\infty\!\!dx\,{x^2\over
    \left(x+m_1^2\right)^2
    \left(x+m_2^2\right)^2
    \left(x+m_L^2\right)^2
    }
  }
After integration by parts, this expression is identical to the
one obtained through the scalar kinetic terms, \drhofinal, as
it should be.
\subsec{Properties of $\delta\rho$}
$\delta\rho$ may be computed and expressed in terms of the
particle masses  $m_1^2,m_2^2,m_L^2$ as 
\eqn\rhom{
  \eqalign{
    \delta\rho
    &=
    {1\over(4\pi v)^2}
    \left[
      \ml+
      {-4\mo\mt\ml+(\mo+\mt)(m_L^4+\mo\mt)\over
        (\mo-\mt)^2}\right.
      \cr&\qquad
      \left.-{2\mo(\mt-\ml)^2(m_1^4-\mt\ml) \log \mo/\ml\over
        (\mo-\mt)^3(\mo-\ml)}
      +{2\mt(\mo-\ml)^2(m_2^4-\mo\ml) \log \mt/\ml\over
        (\mo-\mt)^3(\mt-\ml)}
    \right]\cr
    }
  }
This formula is consistent with the previous results given in
\ref\LIS{C.S. Lim, T. Inami, N. Sakai, \prd{\bf D29} (1984) 1488}.
{}From this expression, it is clear that this {\it is} a
non--decoupling effect since if we scale up all the masses
without changing their ratios, the expression will scale up
quadratically with the mass scale.  This, of course, does not
contradict the decoupling theorem since the coupling constant
cannot be held fixed while all the masses are scaled up this
way.
We also find the following behavior for large $\mo$
\eqn\largeeone{
  \delta\rho=
  {1\over(4\pi v)^2}
  \left[
    \mt+\ml-2{\mt\ml\over\mt-\ml}\log{\mt\over\ml}
    +\c O\left({\mt\over\mo},{\ml\over\mo}\right)
    \right]
  }

The non--decoupling effect arises because an increase in a
particular mass parameter inevitably gives rise to an increase
in the coupling.  The effect is quadratic in the mass, which is
similar to the Yukawa coupling, or to the case dimensionless
coupling \drhosm.  (The Higgs contribution to the $\rho$
parameter is somewhat special since it is suppressed for
symmetry reasons \ref\SCREENING{M.~Veltman, {\sl Acta
    Phys. Pol.}{ \bf B8}(1977); \plb{\bf 70B} (1977) 253\nl
  M.B.~Einhorn, J.~Wudka, \prd{\bf 39D} (1989) 39}.)  If we
analyze the behavior of the non--decoupling effect with respect
to the coupling, it is {\it linear} with respect to the
coupling.
This is a milder dependence than the contribution from a
dimensionless coupling where the behavior is generically
quadratic in the coupling to leading order.  In this sense, we
see that the behavior of the non--decoupling effect due to a
dimensionful coupling lies somewhere in ``between'' that of the
dimensionless coupling and the free case.

\newsec{$S$ parameter}
In this section, we compute the contribution to the so--called
$S$ parameter 
\ref\SREF{B.W.~Lynn, M.~Peskin and R.G.~Stuart in ``Physics at
  LEP'', eds. J.  Ellis and R.D. Peccei (1986)\nl M.~Peskin,
  T.~Takeuchi, \prl{\bf 65} (1990) 964\nl D.C.~Kennedy,
  P.~Langacker \prl{\bf 65} (1990) 2967\nl G.~Altarelli,
  R.~Barbieri \plb{\bf 253B} (1991) 161}
 and the contribution to it measured using the longitudinal
modes of the gauge boson --- the
Nambu--Goldstone bosons --- which we shall call $\tilde S$
\APS. 
\subsec{$\tilde S$}
$\tilde S$ parameter is the amount of mixing between $W^3$ and
$B$ gauge bosons  measured through their longitudinal modes.
It may be calculated within the scalar theory as 
\eqn\sdef{
  \tilde S\equiv-2\pi v^2 \left.{d \over d(p^2)} Z_0
    (p^2)\right|_{p^2=0}
  }
This parameter measures how the effective coupling changes with
the momentum at zero energies and is a physical parameter within
the  scalar theory.

{}From the expression for the scalar self energy in \pifactors, we 
obtain 
\eqn\stint{
  \eqalign{
  \tilde S& =
  \lambda^2 \int_0^\infty\!\!dx\,
  {x\over(x+ m_1^2)^3(x+m_2)}\left[
    {m_1^2\over x+m_1^2}-{2  x m_1^2\over (x+m_1^2)^2}
    \right]\cr
     &=-{1\over12\pi}
    {(\mo-\ml)(\mt-\ml)\left[m_1^6-m_2^6+9(m_1^4 \mt-\mo m_2^4)
        -6\mo\mt (\mo+\mt)\log \mo/\mt\right]\over
      (\mo-\mt)^5} 
    \cr}
}

Expressed this way, the $\tilde S$ parameter is again clearly a
non--decoupling effect.  If we increase the masses while keeping
their ratios fixed, the parameter stays constant, so that this
is again a non--decoupling effect.  $\tilde S$
has a milder behavior with respect to the particle mass when
compared with the $\rho$ parameter.  Especially for large $\mo$,
it has a simple behavior 
\eqn\stlargemone{
  \tilde S={1\over12\pi}{\ml-\mt\over\mo}\left[1
  +\c O\left({\mt\over\mo},{\ml\over\mo}\right)\right]
  }
If we keep the mass ratio between $m_1^2$ and $m_2^2$ and
increase the particle masses, the $\tilde S$ parameter remains
constant.  This is reminiscent of the behavior of the
contribution to the $S$ parameter from the Yukawa coupling in
the Standard Model, wherein the part that remains constant with
the particle mass is generated by the longitudinal modes of the
gauge bosons, as it was here.
\subsec{$S$ parameter}
The so--called $S$ parameter \SREF\ may be defined as the amount
of mixing between $W_3$ and $B$ as in
\eqn\sdefw{S\equiv-{16\pi\over gg'}
        {d\over dp^2}\Pi^{W^3B}(0)
      }
This may be computed from the graphs in 
\fig\figsw{}
\nl\nl\centerline{
  \psannotate{\psboxto(0cm;\vsiz){vcurr.eps}}
  {\fillinggrid%
    \at(-1.0\pscm;3.0\pscm){$W^3$}%
    \at(6.5\pscm;3.0\pscm){$B$}%
    \at(5.0\pscm;4.9\pscm){$\tpl,\tpr$}%
    \at(1.0\pscm;0.0\pscm){$\tpl$}%
    }
  \hskip1.2cm
  \psannotate{\psboxto(0cm;\vsiz){vcurr.eps}}
  {\fillinggrid%
    \at(-1.0\pscm;3.0\pscm){$W^3$}%
    \at(6.5\pscm;3.0\pscm){$B$}%
    \at(5.0\pscm;4.9\pscm){$\tpr$}%
    \at(1.0\pscm;0.0\pscm){$\tpr$}%
    }
  \hskip1.2cm
  \psannotate{\psboxto(0cm;\vsiz){vcurr.eps}}
  {\fillinggrid%
    \at(-1.0\pscm;3.0\pscm){$W^3$}%
    \at(6.5\pscm;3.0\pscm){$B$}%
    \at(5.0\pscm;4.9\pscm){$\left(\varphi_L\right)_2$}%
    \at(0.0\pscm;-0.2\pscm){$\left(\varphi_L\right)_2$}%
    }
} \nl
\figsw\ {\it Graphs contributing to  the $S$ parameter}\nl\nl
to obtain
\eqn\pisw{
  \eqalign{
    \Pi^{W^3B}_{\mu\nu} = 
    {gg'\over12}\Bigl[ & 
    \cst(1+3\sst)\kmn\mo\mo
    +\sst(1+3\cst)\kmn\mt\mt
    \cr &\qquad
    -\kmn\ml\ml-6\cst\sst\kmn\mo\mt
    \Bigr]\cr
    }
  }
where
\eqn\kmndef{
  \kmn ab\equiv \dk{(2k+p)_\mu(2k+p)_\nu
    \over
    (k^2+a)\left((k+p)^2+b\right)     }
}
After some computation, we obtain the $S$ parameter as 
\eqn\sparam{
  \eqalign{
    S &=
    {1\over12\pi}\biggl\{
    {(\mo-\ml)(\mt-\ml)      \over(\mo-\mt)^5}
    \biggl[{5\over3}(m_1^6-m_2^6)-9\mo\mt(\mo-\mt)
    \cr&\qquad    
    +\left(-m_1^6-m_2^6 +3\mo\mt(\mo+\mt)\right)\log
    {\mo\over\mt}\biggr]      
    \cr&\qquad    
    -Y_L{-(\mt-\ml)\log{\mo/\ml}+(\mo-\ml)\log\mt/\ml
        \over\mo-\mt}\biggr\}\cr
      }
}
We compare $\tilde S$ and $S$ and obtain the difference as
\eqn\sresult{
  \eqalign{
    S-\tilde S &=
    {1\over12\pi}\biggl\{
    {(\mo-\ml)(\mt-\ml)
    \left[8/3(m_1^6-m_2^6)
        +\left(\mo+\mt\right)^3\log {\mo/\mt}\right]            
      \over(\mo-\mt)^5}
    \cr&\qquad
    -Y_L{-(\mt-\ml)\log{\mo/\ml}+(\mo-\ml)\log\mt/\ml
        \over\mo-\mt}\biggr\}\cr
      }
  }
The difference, unlike the case of the Yukawa coupling for the
fermions, is {\it not} proportional to $Y_L$ but rather to both
$Y_{L,R}$.  For large $\mo$, the difference in proportional to
the U$(1)_Y$ hypercharge $Y_L$, in a manner reminiscent of the
Yukawa coupling case.
\eqn\largeone{
  S-\tilde S =
  -{1\over12\pi}Y_L\log{\mt\over\ml}
  +\c O\left({\mt\over\mo},{\ml\over\mo}\right)
}

In the case of the $\rho$ parameter, the longitudinal
contribution and the full contribution were identical, as is
required by the gauge symmetry.  For the $S$ parameter we find
that it is not so
\PP\FMM;
the difference may be understood as
follows.  Considering the effect from the effective action, the
lowest dimension operator that contributes to the $S$ parameter
has dimension eight.  At dimension eight, only one operator,
$\left(\phi^\dagger D_\mu D_\nu \phi\right)^2$ contributes to
the $\tilde S$ parameter.  There is an additional operator
$\left(\phi^\dagger W_{\mu\nu}\phi\right)
\left(\phi^\dagger B_{\mu\nu}\phi\right)$ that contributes to
$S$, that does {\it not} involve the scalar kinetic term.  It
does not seem possible to determine this coefficient within the
scalar theory.
\newsec{Discussion}
In this note, we showed how the cubic coupling in the scalar
theory may give rise to non--decoupling effects perturbatively
and analyzed these effects, such as their mass dependence.
It is widely known that the dimensionless couplings can give
rise to non--decoupling effects and the dimensionless nature of
the couplings have sometimes been stressed.  It is on the other
hand clear that the mass (quadratic) terms by themselves do not
give rise to non--decoupling effects. We have clearly shown that
the dimensionful scalar cubic coupling does give rise to
non--decoupling effects by itself.  This basically completes the
picture as to which renormalizable couplings may give rise to
non--decoupling effects and which do not.

The contribution to the $\rho$ and the $S$ parameters from the
scalar trilinear coupling, somewhat surprisingly, turn out to be
substantially more complicated than those for the Yukawa
coupling or the Higgs coupling.  When reinterpreted in terms of
the coupling constant, the $\delta\rho$ contribution is linear
in the coupling to leading order.  This is a feature not
observed in the non--decoupling effects due to dimensionless
couplings.  Roughly speaking, the strength of the contribution
is somewhat weaker than that from a dimensionless coupling.

The gauge bosons derive their masses from their couplings to the
Nambu--Goldstone bosons.  Due to gauge invariance, one might
expect that the non--decoupling effects might be obtained by
examining just the Nambu--Goldstone dynamics.  If this were
possible, it would be instrumental in analyzing the systematics
of these effects, since we may summarize the contributions within
the scalar theory.  Also, if we only had to perform scalar
computations, the work would be facilitated substantially.
This is somewhat similar in philosophy to the so--called
``equivalence theorems'' used in high energy scattering of $W$
bosons. \ref\EQUIV{
  J.S. Bell, \npb{\bf B60} (1973) 428\nl
  C.H. Lewellyn Smith, \plb{\bf 46B} (1973) 233\nl
  J.M. Cornwall, D.N. Levin, G. Tiktopoulos, \prl{\bf 30} (1973) 
  1268\nl
  M. Chanowitz, M.K. Gaillard, \plb{\bf 142B} (1984) 85}
For leading contributions to the $\rho$ parameter, this is
indeed the case, as was shown some time ago \LYTEL.  In general,
this clearly can {\it not } be true since the scalar theory
knows nothing about the U$(1)_Y$ hypercharge.  Even if we were
to ignore the hypercharge, the effect still needs not be
computable by examining only the longitudinal modes in general
\PP.  We indeed find that the contribution to the $S$ parameter
is different from the contribution obtained from the
longitudinal modes of the gauge bosons for which we gave reasons
as to why this is so.
In our example, the difference is {\it not} proportional to
$Y_L$ even at leading order, which seems to be a rather novel
feature. 

Finally, it may be prudent to reflect on what the
non--decoupling effects are in a precise sense: In some
theories, a mass parameter can {\it not} be generated by a
quadratic mass term in the Lagrangian for symmetry reasons.  In
such cases, it may be possible to generate the mass term through
a coupling constant and a vacuum expectation value of a field.
If such cases, the increase in the mass parameter necessarily
entails stronger coupling and we evade the assumptions of the
decoupling theorem.  This {\it may} lead to quantum effects that
do not decrease with the increase in this mass parameter while
this is {\it not} inevitable.  In the Standard Model, the
fermion and the Higgs mass terms are forbidden due to gauge
symmetry.  In the case we analyzed, the symmetry reasons ---
which can be gauge symmetry, just as in the Standard Model ---
forbid the appearance of certain mass parameters, more
specifically, the mixing between $(\varphi_L)_1$ and
$\varphi_R$.  This mixing will necessarily give rise to certain
non--decoupling effects.
\bigskip\noindent{\bf Acknowledgments:} We would like to thank
Santiago Peris for invaluable discussions.  This work was
supported in part by the Grant--in--Aid from the Ministry of
Education, Science, Sports and Culture and grants from Keio
University.
\listrefs\end